\documentclass[preprint]{revtex4-1}

\usepackage{amsmath}
\usepackage{amssymb}
\usepackage{graphics}
\usepackage{graphicx}
\usepackage{url}

\newcommand{\eqnumref}[1]{(\ref{eq:#1})}
\renewcommand{\eqref}[1]{eq.~\eqnumref{#1}}

\newcommand{\figref}[1]{Fig.~\ref{fig:#1}}
\newcommand{\Figref}[1]{Figure~\ref{fig:#1}}
\newcommand{\PUNT}[1]{}

\newcommand{\citeasnoun}[1]{Ref.~\citenum{#1}}

\begin{document}

\title{Tandem photonic-crystal thin films surpassing Lambertian
  \\ light-trapping limit over broad bandwidth and angular range}

\author{Ardavan Oskooi}
\email{oskooi@qoe.kuee.kyoto-u.ac.jp}

\author{Yoshinori Tanaka}

\author{Susumu Noda}

\affiliation{Department of Electronic Science \& Engineering, Kyoto
  University, Kyoto, 615-8510, Japan}

\begin{abstract}
Random surface texturing of an optically-thick film to increase the
path length of scattered light rays, first proposed nearly thirty
years ago, has thus far remained the most effective approach for
photon absorption over the widest set of conditions. Here using recent
advances in computational electrodynamics we describe a general
strategy for the design of a silicon thin film applicable to
photovoltaic cells based on a quasi-resonant approach to light
trapping where two partially-disordered photonic-crystal slabs,
stacked vertically on top of each other, have large absorption that
surpasses the Lambertian limit over a broad bandwidth and angular
range.
\end{abstract}


\maketitle

\begin{figure}[t]
{\centering
  \includegraphics[width=1.0\columnwidth]{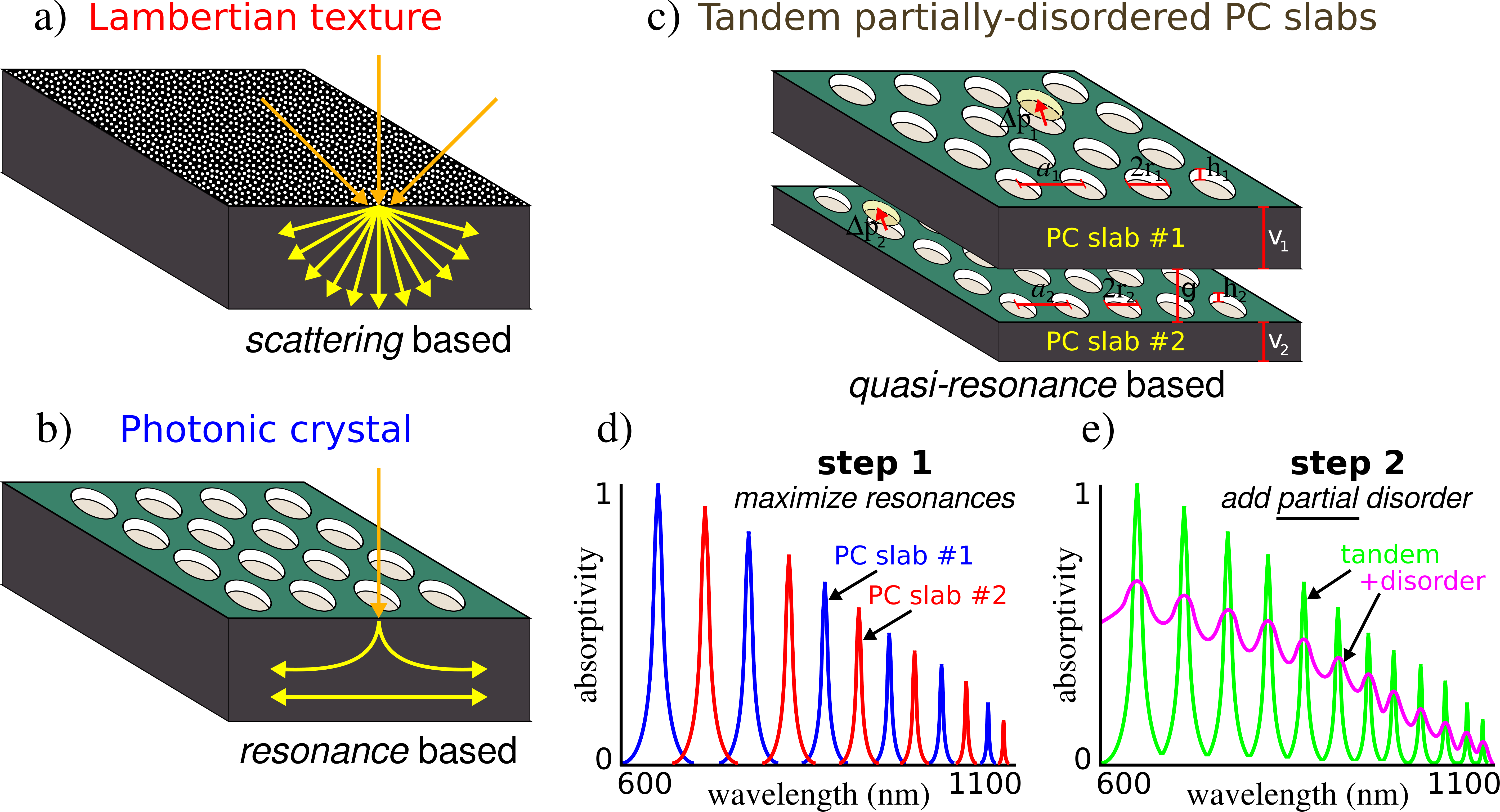} \par}
\caption{(a) Random or so-called Lambertian texturing of the surface
  to isotropically scatter incident light rays into the plane of a
  weakly-absorbing film so as to increase the optical path
  length. First proposed nearly thirty years ago, this has thus far
  remained the most effective approach for light trapping over the
  widest set of conditions. (b) Photonic-crystal slab and other
  nanostructured designs in which light trapping occurs by resonant
  absorption into a guided mode depend on delicate wave-interference
  effects and are thus intrinsically narrowband and restricted to a
  small angular cone. (c) Tandem arrangement of two PC slabs, both
  consisting of a square lattice of holes in silicon, stacked
  vertically on top of each other. Shown are the degrees of freedom --
  slab thicknesses v$_1$ and v$_2$, lattice parameters $a_1$ and
  $a_2$, hole radii r$_1$ and r$_2$, hole heights h$_1$ and h$_2$ and
  gap separation $g$ -- used in the topology optimization to (d)
  engineer as many non-overlapping resonances over the wideband solar
  spectrum as possible. Following this, (e) each hole is perturbed
  from its position in the unperturbed lattice by amounts $\Delta p_1$
  ($\Delta p_2$) chosen randomly from a uniform distribution of values
  between 0 and $\overline{\Delta p_1}$ ($\overline{\Delta p_2}$) for
  both orthogonal in-plane directions of the top (bottom) slab to
  boost light trapping and robustness by creating a more-uniform
  absorption profile.}
 \label{fig:design_approach}
\end{figure}

\noindent One of the fundamental issues underlying the design of
silicon photovoltaic (PV) cells for use in realistic settings is the
maximum absorption of incident sunlight over the widest possible range
of wavelengths, polarizations and angles using the thinnest material
possible. While random or so-called Lambertian texturing of the
surface to isotropically scatter incident light rays into a
weakly-absorbing thick film so as to increase the optical path length,
as shown in ~\figref{design_approach}(a), has thus far remained the
most effective approach for light trapping over a wideband
spectrum~\cite{Yablonovitch82b,Yablonovitch82}, recent thin-film
nanostructured designs including photonic
crystals~\cite{Joannopoulos08} shown in ~\figref{design_approach}(b)
exploiting the more-complicated wave effects of photons have explored
the possibility of superior
performance~\cite{Zhou08,Chutinan09,Park09,Garnett10,Mallick10,Zhu10,Han10,Yu10,Fahr11,Sheng11,Bozzola12,Wang12,Biswas12,Munday12,Martins12}
but have been mainly limited to narrow bandwidths, select
polarizations or a restricted angular cone typical of delicate
resonance-based phenomenon. The introduction of strong disorder, while
improving robustness, nevertheless comes at the expense of light
trapping relative to the unperturbed case~\cite{Oskooi12,Vynck12}. As
a result, no proposal for a nanostructured silicon thin film capable
of robust, super-Lambertian absorption over a large fraction of the
solar spectrum has yet been made. In this work, we present a new
approach to light trapping made possible by recent advances in
computational electrodynamics~\cite{Taflove13} based on the
quasi-resonant absorption of photons that combines the large
absorption of optical resonances with the broadband and robust
characteristics of disordered systems via a stacked arrangement of
ordered PC slabs augmented with partial disorder that surpasses by a
wide margin the performance of an idealized Lambertian scatterer over
a broad spectrum and angular cone.

Our tandem design, consisting of the same silicon film structured in
two different ways and stacked vertically as shown in
~\figref{design_approach}(c), is the photonic analogue of the
multi-junction cell that employs three or more \emph{different}
semiconductor films where the electronic bandgaps add complimentarily
to obtain wideband absorption. Here we demonstrate the utility of a
photonic approach, employing geometric structure alone, to enhance
light trapping which also offers improved performance but does not
incur the constraints and limitations of optimally combining multiple
material-specific electronic bandgaps as well as the significant
fabrication challenges and costs of epitaxially growing films with
mismatched atomic lattices on top of one another. We outline a
two-part design strategy based on first maximizing, with a
few-parameter gradient-free topology optimization the number of
resonant-absorption modes by using two crystalline-silicon PC slabs
with a fixed total thickness of 1$\mu$m stacked on top of each other
such that their individual resonances add complimentarily over the
wideband spectrum; and then in the final step introducing a partial
amount of disorder to both lattices to maximize the overall light
trapping and boost robustness to go well beyond the Lambertian limit.

In our earlier work, we showed how individual resonant-absorption
peaks of a thin-film PC slab can be broadened using partial disorder
leading to an overall enhancement of the wideband absorption spectra
~\cite{Oskooi12}. To understand quantitatively why disorder increases
broadband light trapping in a PC, we use coupled-mode theory to derive
an analytical expression for a single absorption resonance (at a
frequency of $\omega_0$) which has only one coupling channel for
external light (a slight simplification which helps to make clear the
role of disorder) in terms of the decay lifetimes (proportional to the
quality factor) for radiation ($\tau_{rad}$) and absorption
($\tau_{abs}$) by the material~\cite{Joannopoulos08}:

\begin{equation}
A(\omega)=\frac{\frac{4}{\tau_{rad}\tau_{abs}}}{(\omega-\omega_0)^2+\left(\frac{1}{\tau_{rad}}+\frac{1}{\tau_{abs}}\right)^2}.
\label{eq:single_peak}
\end{equation}

Broadband absorption for a thin-film PC design, consisting of a
collection of such individual Lorentzian peaks, necessitates that we
consider the \emph{total} area spanned by ~\eqref{single_peak} which
is equivalent to its absorption cross section:

\begin{equation}
\int_{-\infty}^{\infty}A(\omega)d\omega=\frac{4\pi}{\tau_{rad}+\tau_{abs}}.
\label{eq:peak_area}
\end{equation}

The effect of disorder is to reduce the peak height but more
importantly to broaden the peak width (proportional to
1/$\tau_{rad}$+1/$\tau_{abs}$) via primarily a \emph{decrease} in
$\tau_{rad}$ which therefore leads to an overall \emph{increase} in
broadband absorption from ~\eqref{peak_area} (though $\tau_{abs}$ also
changes with disorder due to variations in the nature of the guided
mode, the change is much less pronounced than $\tau_{rad}$ mainly
because the material absorption coefficient is fixed). Note that this
analysis is only valid when coupling to a \emph{resonant} Bloch mode
which is why introducing too much disorder and eliminating the peaks
altogether, thus transitioning to \emph{non-resonant}
Anderson-localized modes, results in sub-optimal light
trapping~\cite{Oskooi12}.

We consider the absorption of solar radiation in the wavelength regime
spanning 600nm to 1100nm in which silicon is poorly absorbing and thus
weak coupling to resonant Bloch modes of the PC most apparent. The
overall light-trapping efficiency of each design can be quantified
relative to an ideal perfect absorber which has unity absorptivity
over the wavelength interval of interest by assuming that each
absorbed photon with energy greater than the bandgap of silicon
generates an exciton which contributes directly to the short-circuit
current (this is equivalent to an internal quantum efficiency of
100\%). This corresponds to the following definition of light-trapping
efficiency:

\begin{equation}
\frac{\int_{600nm}^{1100nm}\lambda \mathcal{I}(\lambda)\mathcal{A}(\lambda)d\lambda}{\int_{600nm}^{1100nm}\lambda \mathcal{I}(\lambda)d\lambda},
\label{eq:efficiency}
\end{equation}

where $\mathcal{I}(\lambda)$ is the terrestrial power density per unit
wavelength from the sun at AM1.5~\cite{ASTM05} and
$\mathcal{A}(\lambda)$ is the absorptivity of the film.

The design strategy of maximizing the light-trapping efficiency by
controllably introducing a partial amount of disorder to obtain just
the right dose of peak broadening ultimately results in a more-uniform
absorption profile where the absorptivity in the inter-peak regions is
increased at the expense of the height of all peaks. This therefore
suggests that in order to most effectively make use of partial
disorder for light-trapping enhancement in a thin film the number of
resonant modes must first be made \emph{as large as possible} so that
there is little bandwidth separation between peaks: by extending our
previous slab design to include not one but \emph{two} PC lattices
stacked on top of one other and separated by a non-absorbing nanoscale
gap layer, such that the absorption spectra of each lattice when
combined adds complimentarily (i.e., regions of low absorption in one
lattice are compensated for by the high absorption of the other), the
resonant-absorption characteristics of the PC augmented by partial
disorder can potentially be exploited to the fullest extent possible
for broadband absorption while also remaining feasible to large-scale
industrial manufacturing. The low-index gap separation layer itself
also provides additional mechanisms for light localization in
nanostructured media that further contributes to enhancing absorption
in the adjacent high-index layers while its effect on scattering-based
Lambertian-textured films is marginal. A schematic of this design
approach, somewhat exaggerated for illustrative purposes, is shown in
\figref{design_approach}(d) and (e) in which the tandem structure is
first optimized for \emph{peak density} in the spectrum (which amounts
to maximizing the number of non-overlapping resonant absorption modes
of the two constituent PC lattices) and subsequently these narrow
closely-spaced resonances are slightly broadened with the addition of
disorder to create a more-uniform absorption profile that is large in
amplitude, broadband and robust to incident radiation conditions. We
consider here for simplicity the case of two PC slabs in vacuum
separated by an air gap which both incorporates all essential physical
phenomena and has direct applications to conventional thin-film PV
cells in which individual layers including both high-index
semiconductors and low-index transparent conductive oxides are grown
by thin-film deposition tools~\cite{Brendel03,Poortmans06}. There is
no need in the scope of the present work where the focus is solely on
photon absorption in silicon to include an anti-reflection (AR)
coating in the front or a perfect reflector in the back as would be
customary in an actual PV cell since the role of both components is
mainly to enhance, oftentimes significantly~\cite{Bermel07,Mallick10},
the absorption of \emph{existing} resonances in the nanostructured
films but not to give rise to new ones. Due to the complimentary way
that the individual peaks of the two PCs combine in the tandem
structure, a simple square lattice arrangement is adequate for good
performance rendering unnecessary the need for intricate superlattice
structures~\cite{Yu10,Martins12}.

To perform the topology optimization, we combine the capabilities of
Meep, a freely-available open-source finite-difference time-domain
(FDTD) tool~\cite{Oskooi10}, to compute the absorption spectra at
normal incidence with the nonlinear-optimization routines of
NLopt~\cite{NLopt} (details in Supplementary Information). Here we
need not consider absorption at off-normal incidence since the
addition of disorder in the final step will automatically reduce the
sensitivity to incident radiation conditions~\cite{Oskooi12}. Accurate
topology optimization is made possible using Meep's subpixel
averaging~\cite{Farjadpour06,Oskooi10,Taflove13} which also
significantly reduces the size of the computation by lowering the
minimum spatial resolution required for reliable results. This is
important since the objective function is iterated over a large number
of times to explore small changes to geometrical parameters in order
to engineer as many absorption resonances over the wideband spectrum
as possible which tend to be highly susceptible to numerical artifacts
introduced by ``staircased''
features~\cite{Farjadpour06,Oskooi10,Taflove13} enabling FDTD to be
used as a versatile 3D design optimization tool. We use intrinsic
crystalline silicon as our absorbing material and incorporate its full
broadband complex refractive index profile~\cite{Green08} into the
FDTD simulations with accuracy even near its indirect bandgap of
1108nm where absorption is almost negligible to obtain
experimentally-realistic results (see Supplementary Information for
more details). Although, for generality, modeling a tandem structure
consisting of two completely-independent PC slabs with arbitrary
lattice constants is preferred, incorporating two distinct unit cells
into a single 3D simulation is computationally impractical yet this is
a minor design limitation as the other six structural parameters -- as
shown in \figref{design_approach}(c): the thickness of the top Si
layer $v_t$ (the bottom thickness $v_b$ is known since the total
thickness is fixed at 1$\mu$m), the gap thickness $g$, the radius and
height of the holes in the top and bottom lattices $r_t$, $h_t$,
$r_b$, $h_b$ -- provide sufficient flexibility for creating
out-coupled Bloch-mode resonances over the entire range of the
broadband solar spectrum. We also investigated the computationally
tractable case of two PC slabs with lattice constants differing by a
factor of two though the results were not found to be an improvement
to the single lattice-constant design. A planewave source is incident
from above onto the tandem structure and the absorption spectrum
$\mathcal{A}(\lambda)$, equivalent to 1-reflection-transmission, is
calculated by Fourier-transforming the response to a short pulse. The
absorptivity threshold used by the objective function to count the
total number of peaks in the spectrum is taken to be that of our
baseline performance metric: an equivalent 1$\mu$m-thick Si film with
Lambertian-textured surfaces~\cite{Yablonovitch82,Deckman83} which has
an efficiency of 43.0\% in our wavelength interval (computed using the
same fitted material parameters as used in the simulations). Since
resonances with especially-large peak amplitudes contribute most to
increasing the overall efficiency when broadened, we add an extra 30\%
to our absorptivity peak threshold at each wavelength which while
making the problem more challenging gives rise to better results. We
impose no restrictions on the peak width or spacing relative to other
peaks although these could potentially be used for further
refinement. Once an optimal set of parameters for the two-lattice
structure is determined by running multiple times with different
randomly-chosen initial values to explore various local optima, we
then form a supercell consisting of 10x10 unit cells of the optimal
structure and add positional disorder to each unperturbed hole in both
lattices (while ensuring no overlap between holes to conserve the
filling fraction) by an amount $\Delta p_1$ ($\Delta p_2$) chosen
randomly from a uniform distribution of values between 0 and
$\overline{\Delta p_1}$ ($\overline{\Delta p_2}$) for both orthogonal
in-plane directions for the top (bottom) slab. Three separate
simulations are made for each structure and the results averaged due
to the random nature of the design.

\begin{figure}[t]
{\centering
  \includegraphics[width=1.0\columnwidth]{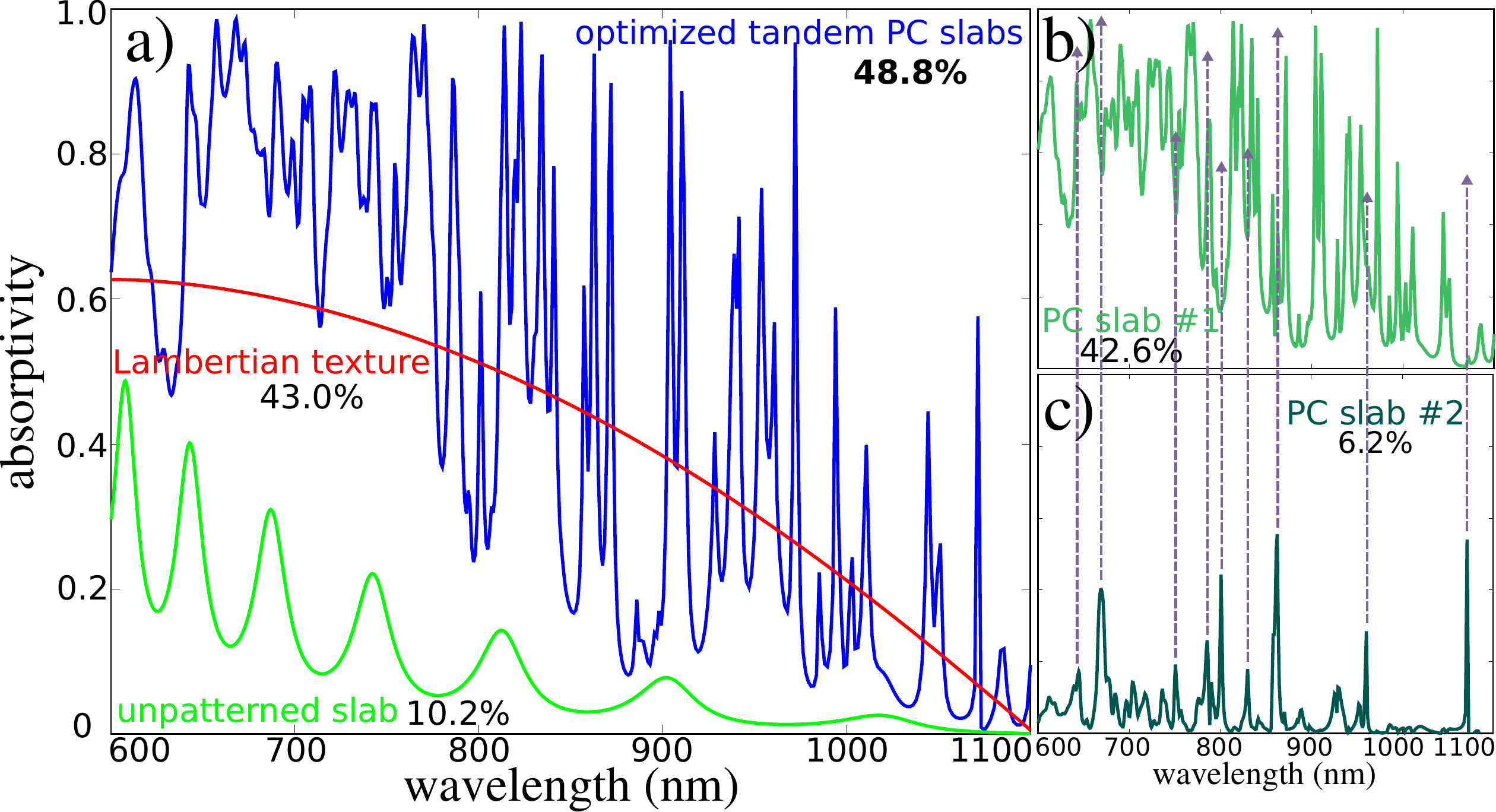} \par}
\caption{(a) Absorption versus wavelength profile at normal incidence
  for three thin-film PV designs each with a total crystalline-silicon
  thickness of 1$\mu$m: an unpatterned slab (green), a
  Lambertian-textured slab (red) [obtained from eq. (1) of
    ~\citeasnoun{Deckman83}] and the topology-optimized tandem PC
  slabs (blue). The tandem PC slabs both consist of a square lattice
  (periodicity, $a$=641nm) of holes in silicon separated by a 228nm
  gap. Shown for each design is the photon-absorption efficiency
  defined in ~\eqref{efficiency} which is a measure of light trapping
  relative to a perfect absorber. (b) and (c) Individual absorption
  spectra for the top (slab \#1: $v_t$=708nm, $r_t$=236nm,
  $h_t$=260nm) and bottom (slab \#2: $v_b$=292nm, $r_b$=199nm,
  $h_b$=244nm) PC slabs of the optimized tandem design demonstrating
  how the resonances add complimentarily over the broadband spectrum.}
 \label{fig:tandem_a641}
\end{figure}

\Figref{tandem_a641}(a) shows the absorptivity spectra for three
thin-film designs each with a total crystalline-silicon thickness of
1$\mu$m: an unpatterned film, a Lambertian-textured film [obtained
  from eq. (1) of ~\citeasnoun{Deckman83}] and finally the
topology-optimized tandem design consisting of two PC slabs (top:
thickness 708nm, hole radius 236nm, hole height 260nm and bottom:
thickness 292nm, hole radius 199nm, hole height 244nm) with a lattice
parameter of 641nm separated by a 228nm gap. The tandem design has
numerous narrow, high-amplitude peaks -- signatures of the
coherent-resonant Bloch modes -- that span the entire broadband
spectrum whereas the unpatterned slab has broad Fabry-P\'{e}rot
resonances with low amplitude. The complimentary way that the
resonances of the individual slabs combine to form the tandem
structure can be seen in \Figref{tandem_a641}(b) and (c) where the top
slab accounts for most of the total number of peaks while the bottom
slab contributes a few key resonances particularly at longer
wavelengths. Note that while the Lambertian-textured slab has little
and diminishing absorption at long wavelengths where the absorption
coefficient of crystalline silicon is vanishingly small the PC design,
due to its resonant nature, has large absorption albeit appearing only
as very-narrow peaks (due to the rate-matching phenomena discussed
previously that underlies the resonant coupling between radiation and
absorption by the material, as silicon's absorption coefficient
becomes smaller at larger wavelengths leading to a corresponding
increase in $\tau_{abs}$~\cite{Joannopoulos08}, the total lifetime of
the resonant mode $\tau_{tot}$ being
1/$\tau_{tot}$=1/$\tau_{abs}$+1/$\tau_{rad}$ also increases resulting
in the inversely-proportional peak width generally becoming narrower
which can be seen in ~\figref{tandem_a641}). Nevertheless, the tandem
design with its maximized peak density, outperforms the Lambertian
texture in light-trapping efficiency (48.8\% versus 43.0\%) although
at off-normal angles and different polarizations of incident light the
unperturbed lattices' resonance-based performance degrades
significantly to below the Lambertian limit. Within the stacked
arrangement, the efficiencies of the top and bottom slabs are 42.6\%
and 6.2\%, respectively, while in isolation they are 37.9\% and 22.3\%
highlighting in part the importance of inter-slab interactions to the
overall absorption of the tandem design. For comparison, the optimized
1$\mu$m-thick single-slab design (lattice parameter, hole radius and
height of 640nm, 256nm and 400nm, respectively) produces seven fewer
resonances than the tandem design over the same wideband spectrum and
therefore had a lower efficiency: nearly 5\% below in absolute terms,
yet still above the Lambertian limit.

\begin{figure}[t]
{\centering
  \includegraphics[width=1.0\columnwidth]{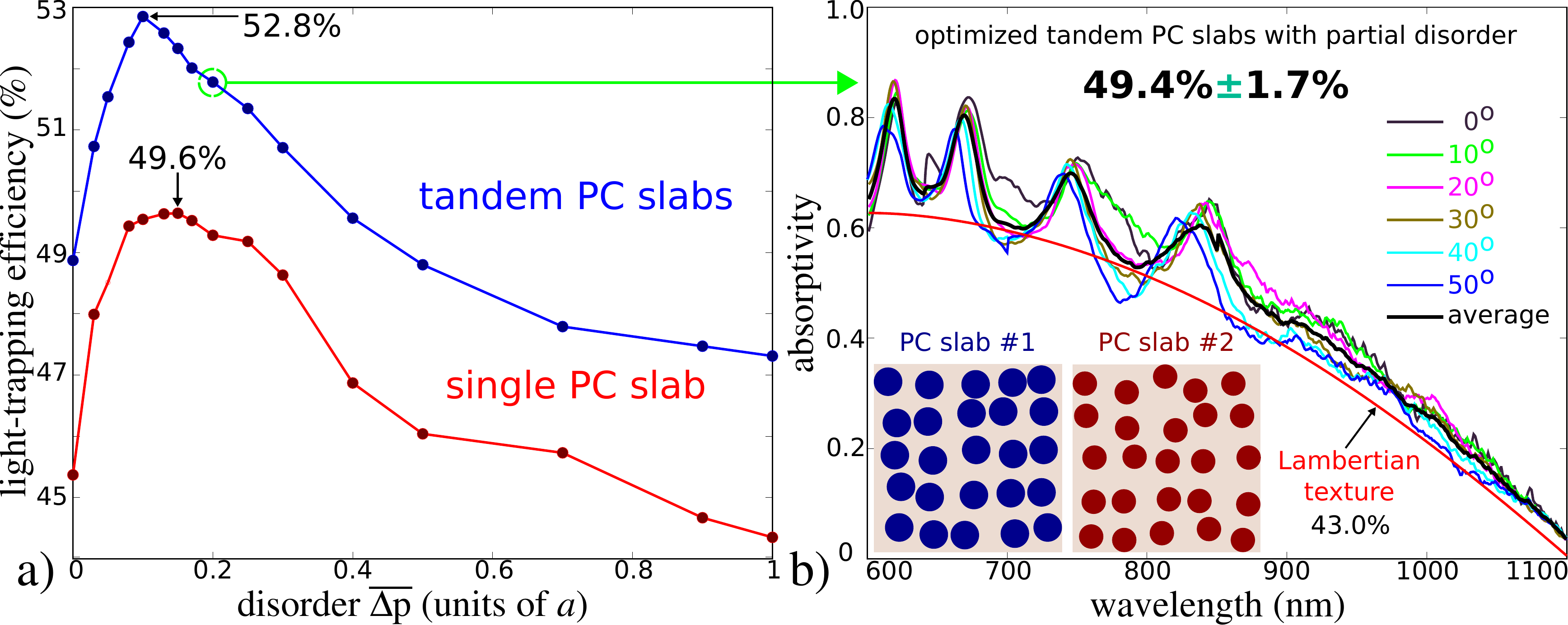} \par}
\caption{(a) Light-trapping efficiency from ~\eqref{efficiency} as
  computed from the absorption profile at normal incidence versus hole
  positional disorder for the topology-optimized tandem slabs (blue)
  and single slab (red) showing that partial disorder (tandem:
  $\overline{\Delta p_1}$=$\overline{\Delta p_2}$=0.1$a$, single:
  $\overline{\Delta p}$=0.15$a$) maximizes the light trapping (tandem:
  52.8\%, single: 49.6\%) while additional disorder is sub optimal and
  leads to a steady decline. Note that the tandem design is nearly
  twice as effective as the single-slab design in surpassing the
  Lambertian limit. (b) Absorption versus wavelength profile at normal
  (0$^{\circ}$) incidence and five off-normal (10$^{\circ}$,
  20$^{\circ}$, 30$^{\circ}$, 40$^{\circ}$, 50$^{\circ}$) angles of
  incidence averaged over both $\mathcal{S}$ and $\mathcal{P}$
  polarizations of the optimized tandem design with the addition of
  partial disorder of $\overline{\Delta p_1}$=0.2$a$ and
  $\overline{\Delta p_2}$=0.25$a$ as shown in the inset
  schematics. The absorption profile for the individual angles are
  colored while the average of all the data is shown in black which
  exceeds the Lambertian texture over the entire interval resulting in
  an overall light-trapping efficiency that is approximately 10\%
  greater.}
\label{fig:tandem_random}
\end{figure}

By proceeding to controllably introduce a partial amount of disorder
into the topology-optimized tandem design, we can simultaneously boost
efficiency and improve robustness to exceed the Lambertian limit by an
even wider margin over a large angular
cone. ~\figref{tandem_random}(a) is a plot of the efficiency from
~\eqref{efficiency} versus disorder for the optimized tandem-slabs and
single-slab designs at normal incidence and shows that a positional
disorder of approximately $\overline{\Delta p_1}$=$\overline{\Delta
  p_2}$=0.1$a$ for the tandem slabs and $\overline{\Delta p}$=0.15$a$
for the single slab results in maximal light trapping of 9.8\% and
6.6\% above the Lambertian limit, respectively, while additional
disorder beyond these partial amounts leads to a steady decrease of
the efficiency in line with the analysis presented earlier. The tandem
design therefore is roughly twice as effective as the single slab in
overcoming the Lambertian limit due mostly to facilitating a larger
number of absorption resonances. We quantify the performance
robustness of each design as the standard deviation of the efficiency
averaged over normal (0$^{\circ}$) incidence and five off-normal
(10$^{\circ}$, 20$^{\circ}$, 30$^{\circ}$, 40$^{\circ}$, 50$^{\circ}$)
angles of incidence for both $\mathcal{S}$ and $\mathcal{P}$
polarizations. A demonstration involving a larger angular range is
possible but the necessary simulation times to ensure that the Fourier
transforms used to compute the flux spectra have properly converged
become prohibitively long. Since more disorder results in better
robustness~\cite{Oskooi12} which is a key requirement of a practical
solar cell we make a slight trade off and apply not the quantities
which maximize efficiency at normal incidence in the tandem design but
values slightly greater ($\overline{\Delta p_1}$=0.2$a$ and
$\overline{\Delta p_2}$=0.25$a$) where the robustness is substantially
larger: 49.5\%$\pm$2.3\% for the former versus 49.4\%$\pm$1.7\% for
the latter. ~\figref{tandem_random}(b) demonstrates that the average
peformance of this tandem design has greater absorption than the
Lambertian texture at every wavelength resulting in a light-trapping
improvement of almost 10\% above the Lambertian limit.

In summary, we have described a general design strategy derived from a
new conceptual framework of photon capture for a nanostructured
silicon thin film based on the quasi-resonant absorption of photons in
a tandem arrangement of partially-disordered photonic-crystal slabs
separated by a nanoscale gap where the overall light trapping
surpasses a Lambertian-textured film by a wide margin over a large
fraction of the solar spectrum and a broad angular cone.

\section*{Acknowledgments}

This work was supported by Core Research for Evolutional Science and
Technology (CREST) from the Japan Science and Technology Agency. A.O.
was supported by a postdoctoral fellowship from the Japan Society for
the Promotion of Science (JSPS).

\section*{Author Contributions}

A.O. conceived of the entire idea and performed all the simulations
and analysis. A.O. discussed the results and wrote the manuscript with
S.N.  and Y.T. S.N. organized the project and nurtured the environment
for inquiry into how to increase absorption in thin films.

\newpage

\newpage

\newpage

\end{document}